\documentclass[a4paper]{article}

\usepackage{INTERSPEECH2021}
\usepackage{multirow}
\usepackage{cite}
\usepackage{hyperref}
\makeatletter
\newcommand{\thickhline}{%
	\noalign {\ifnum 0=`}\fi \hrule height 1.0pt
	\futurelet \reserved@a \@xhline
}

\title{Unsupervised Domain Adaptation for Dysarthric Speech Detection via \\Domain Adversarial Training and Mutual Information Minimization}
\name{Disong Wang$^{1*}$\thanks{*Work done during internship at Huawei Noah’s Ark Lab}, Liqun Deng$^2$, Yu Ting Yeung$^2$, Xiao Chen$^2$, Xunying Liu$^1$, Helen Meng$^1$}
\address{
  $^1$The Chinese University of Hong Kong, Hong Kong SAR, China\\
  $^2$Huawei Noah’s Ark Lab}
\email{\footnotesize \{dswang,xyliu,hmmeng\}@se.cuhk.edu.hk, \{dengliqun.deng,yeung.yu.ting,chen.xiao2\}@huawei.com}

\begin{document}

\maketitle
\begin{abstract}
  Dysarthric speech detection (DSD) systems aim to detect characteristics of the neuromotor disorder from speech.  Such systems are particularly susceptible to domain mismatch where the training and testing data come from the source and target domains respectively, but the two domains may differ in terms of speech stimuli, disease etiology, etc. It is hard to acquire labelled data in the target domain, due to high costs of annotating sizeable datasets. This paper makes a first attempt to formulate cross-domain DSD as an unsupervised domain adaptation (UDA) problem.  We use labelled source-domain data and unlabelled target-domain data, and propose a multi-task learning strategy, including dysarthria presence classification (DPC), domain adversarial training (DAT) and mutual information minimization (MIM), which aim to learn dysarthria-discriminative and domain-invariant biomarker embeddings.  Specifically, DPC helps biomarker embeddings capture critical indicators of dysarthria; DAT forces biomarker embeddings to be indistinguishable in source and target domains; and MIM further reduces the correlation between biomarker embeddings and domain-related cues. By treating the UASPEECH and TORGO corpora respectively as the source and target domains, experiments show that the incorporation of UDA attains absolute increases of 22.2\% and 20.0\% respectively in utterance-level weighted average recall and speaker-level accuracy.

\end{abstract}

\section{Introduction}

Dysarthria encapsulates various speech disorders caused by a set of neurodegenerative conditions and diseases, such as cerebral palsy, Parkinson’s disease or amyotrophic lateral sclerosis, which lead to poor control of muscles including lips, tongue, jaw, velum and throat \cite{yunusova2008articulatory}. Therefore, patients with dysarthria often produce harsh and breathy speech with unstable prosody and imprecise articulation. To facilitate the clinical diagnosis and treatment of neurological diseases, early onset detection of dysarthric speech may serve as a promising tool. 

Current research on dysarthric speech detection (DSD) mainly focuses on building models trained and validated on the data from the same domain, where high DSD accuracy can be achieved. However, DSD models are less robust under domain mismatch conditions, i.e., DSD models trained by the data from the source domain will suffer from marked performance degradation when they are exposed to data from an unseen target domain with different statistical distributions. The difference may be the types of stimuli, phonetic context, vocal quality, disease etiology, recording environments and devices, etc. Training on labelled data from the target domain will improve DSD accuracy. However, such data is often too costly to acquire \cite{paja2012automated}. Therefore, leveraging available, labelled source data along with unlabelled target data to create a DSD model that generalizes well to target domain is desirable, and can be treated as an \textit{unsupervised} domain adaptation (UDA) problem \cite{ganin2015unsupervised}, as no supervision information is available in the target domain.

To alleviate the domain mismatch issues, domain adversarial training (DAT) and mutual information minimization (MIM) are proposed to extract domain-invariant biomarker embeddings that are used to identify the dysarthria for accurate DSD. The UDA framework consists of three learning tasks: The primary task employs a biomarker encoder to extract biomarker embeddings for dysarthria presence classification (DPC). The second task applies DAT to force biomarker embeddings to be indistinguishable in the source and target domains by deceiving a domain discriminator, so that the biomarker embeddings from source and target domains have similar distributions. The last task strives to minimize the mutual information between the biomarker embeddings and the counterpart domain embeddings that are extracted by a domain encoder, which further removes domain cues in biomarker embeddings. The proposed UDA framework facilitates the learning of biomarker embeddings that are invariant across domains while capturing critical information used for dysarthria detection.

This work paves the way towards the under-explored problem of cross-domain DSD that is widely encountered in practice. The main contribution lies in the novel approach by combining DPC, DAT and MIM for cross-domain DSD that is formulated as an UDA problem for the first time. Extensive experiments have been conducted to verify the effectiveness of proposed methods by using different kinds of neural networks.

\section{Related work}

Diagnosis of speech symptoms is commonly used in the identification of dysarthria. Traditional approaches are involved with clinicians or speech-language pathologists conducting a series of subjective listening tests, e.g., Frenchay Dysarthria Assessment \cite{enderby1980frenchay}, which may be affected by  subjectivity among assessors. This motivates researchers to turn to objective evaluation of dysarthria based on building statistical DSD models, which are economical with potentials for remote patient rehabilitation monitoring \cite{gurugubelli2019perceptually}. To develop an efficient DSD system, previous work mostly design handcrafted acoustic features as biomarkers that capture dysarthric patterns, including prosodic, spectral, phonological and glottal features \cite{norel2018detection,narendra2018dysarthric,kodrasi2019super,mayle2019diagnosing, korzekwa2019interpretable}. Besides, automatic feature extraction from raw speech via a learnable frontend is proposed in \cite{millet2019learning}. Though significant progress has been achieved, effectiveness of previous methods requires further verification under cross-domain conditions. A few previous efforts investigate this problem by carefully designing and selecting domain-robust features for cross-language \cite{orozco2016automatic} and cross-dataset \cite{gillespie2017cross} scenarios. In contrast, this paper focuses on automatically extracting dysarthria-discriminative and domain-invariant biomarkers from simple acoustic features, e.g., mel-spectrograms, which is formulated as an UDA problem. 

UDA has been explored in many speech tasks including automatic speech recognition \cite{sun2017unsupervised,woszczyk2020domain}, speech emotion recognition \cite{abdelwahab2018domain} and speaker recognition \cite{wang2018unsupervised}, where DAT \cite{ganin2016domain} is widely used to remove domain variations and project the data of different domains into the same subspace. There is still much room to apply DAT for the DSD task. Besides, inspired by information theory \cite{gierlichs2008mutual}, MIM is proposed to reduce the dependency between biomarker embeddings and domain-related information. The combination of DAT and MIM forces biomarker embeddings to achieve robustness for detection of dysarthria.  
\begin{figure}[t]
  \centering
  \centerline{\includegraphics[width=8cm]{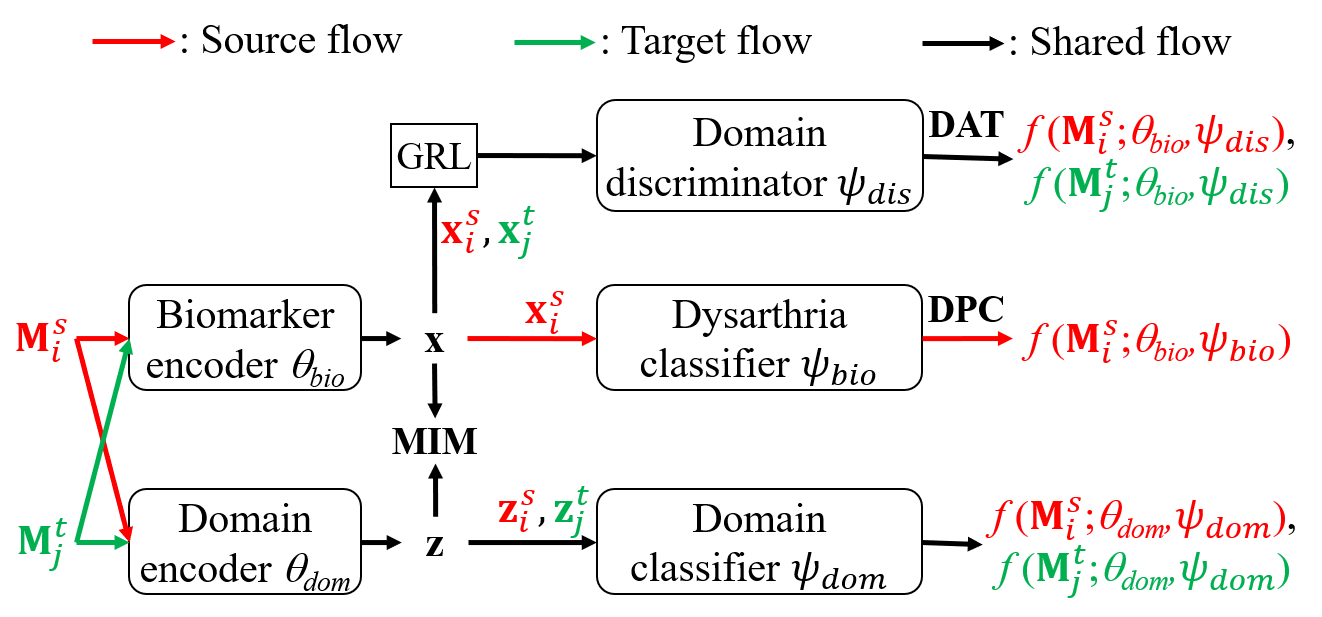}}
  \vspace{-1em}
  \caption{Diagram of the proposed UDA framework for DSD based on multi-task learning, which includes dysarthria presence classification (DPC), domain adversarial training (DAT) and mutual information minimization (MIM). `GRL' denotes the gradient reversal layer.}\label{dia}
  \vspace{-1.8em}
\end{figure}

\vspace{-0.5em}
\section{Proposed approach}
Assuming that there are \textit{I} and \textit{J} utterances with corresponding mel-spectrograms $\{\textbf{M}_{i}^{s}\}$ and $\{\textbf{M}_{j}^{t}\}$ in the source domain and target domain, respectively. Each source mel-spectrogram $\textbf{M}_{i}^{s}$ is associated with a binary label $\textit{y}_{i}^{s}$ denoting whether the corresponding speech is dysarthric, while no such label is provided in the target domain. Given the data $\{\textbf{M}_{i}^{s},\textit{y}_{i}^{s}\}(\textit{i}=1,2,...,\textit{I})$ and $\{\textit{M}_{j}^{t}\} (\textit{j}=1,2,...,\textit{J})$, the goal is to build a DSD system that generalizes well to the target domain. To achieve robustness in a DSD system, we propose a multi-task learning based UDA framework as shown in Figure \ref{dia}, which consists of three learning tasks: dysarthria presence classification, domain adversarial training and mutual information minimization. 

\vspace{-0.6em}
\subsection{Dysarthria presence classification (DPC)}
This primary task performs binary classification for the presence or absence of dysarthria by using the labelled source data. Specifically, a biomarker encoder $\theta_{bio}$ takes in mel-spectrogram $\textbf{M}_{i}^{s}$ to derive a single vector $\textbf{x}_{i}^{s}$, which is denoted as biomarker embedding and fed into the dysarthria classifier $\psi_{bio}$, to give the dysarthria presence posterior $f\left( \textbf{M}_{i}^{s};{\theta_{bio}},{{\psi}_{bio}} \right)$. During training, the biomarker encoder and dysarthria classifier are optimized to minimize the cross-entropy loss as follows:\begin{equation}
\begin{split}
    {{L}_{bio}}=
    -\frac{1}{I}&\sum\nolimits_{i=1}^{I} \big[ y_{i}^{s}\log f\left( \mathbf{M}_{i}^{s};{{\theta }_{bio}},{{\psi }_{bio}} \right)+\\
    &\left( 1-y_{i}^{s} \right)\log \left( 1-f\left( \mathbf{M}_{i}^{s};{{\theta }_{bio}},{{\psi }_{bio}} \right) \right) \big] \label{dys-cls}
\end{split}
\end{equation}

\vspace{-0.8em}
\subsection{Domain adversarial training (DAT)}
The second task applies DAT to render biomarker embeddings indistinguishable in the source and target domains, by introducing a gradient reversal layer (GRL) and a domain discriminator $\psi_{dis}$, as shown in Figure \ref{dia}. During training, parameters of the domain discriminator and biomarker encoder are updated alternatively. On the one hand, by freezing $\theta_{bio}$, the domain discriminator is trained to determine whether the input biomarker embeddings are from the source or target domains by minimizing the discrimination loss \cite{ganin2016domain}:\begin{equation}
\begin{split}
{{L}_{dis}}=&\frac{1}{I}\sum\nolimits_{i=1}^{I}{\log f\left( \textbf{M}_{i}^{s};{{\theta }_{bio}},{{\psi }_{dis}} \right)}+ \\
&\frac{1}{J}\sum\nolimits_{j=1}^{J}{\log \left(1-f\left( \textbf{M}_{j}^{t};{{\theta }_{bio}},{{\psi }_{dis}} \right)\right)}  \label{dis}
\end{split}
\end{equation}On the other hand, by freezing $\psi_{dis}$, the biomarker encoder is trained to maximize the above discrimination loss to deceive the discriminator, which is realized via GRL that passes the data during forward propagation and inverts the sign of the gradient during backward propagation. The alternative processes in training force the domain discriminator and biomarker encoder to compete against each other in an adversarial manner \cite{goodfellow2014generative}, which encourages the distribution of biomarker embeddings across domains to be similar, so that the dysarthria-related cues learned from the source domain in the DPC task remain effective in the target domain. 

\begin{table}[t]
  \label{alg}
  \centering
  \begin{tabular}{ll}
    \toprule[1.5pt]
    \hspace{-0.2cm}\textbf{Algorithm 1.} Training process for UDA based DSD system
    \\
    \midrule
    \hspace{-0.2cm}\textbf{Input:} source data $\{\textbf{M}_{i}^{s},\textit{y}_{i}^{s}\}$, target data $\{\textbf{M}_{j}^{t}\}$, learning rate
    \\
    \hspace{-0.2cm}$\alpha$, $\beta$ and $\gamma$
    \\
    \hspace{-0.2cm}1. \textbf{for} each training iteration \textbf{do}
    \\
    \hspace{-0.2cm}2. \quad freeze $\theta_{bio}$, $\psi_{bio}$, $\theta_{dom}$, $\psi_{dom}$ and $\phi$, compute discrimi- 
    \\
    \hspace{-0.2cm}\quad \quad nation loss (\ref{dis}) using $\{\textbf{M}_{i}^{s}\}$ and $\{\textbf{M}_{j}^{t}\}$, then update $\psi_{dis}$: 
    \\
    \hspace{-0.2cm}3. \qquad \quad $\psi_{dis}\leftarrow \psi_{dis}-\alpha {\nabla}_{\psi_{dis}} L_{dis}$
    \\
    \hspace{-0.2cm}4. \quad freeze $\theta_{bio}$, $\psi_{bio}$, $\theta_{dom}$, $\psi_{dom}$ and $\psi_{dis}$, compute log-
    \\
    \hspace{-0.2cm}\quad \quad likelihood (\ref{lld}) using $\{\textbf{M}_{i}^{s}\}$ and $\{\textbf{M}_{j}^{t}\}$, then update $\phi$:
    \\
    \hspace{-0.2cm}5. \qquad \quad $\phi\leftarrow \phi+\beta {\nabla}_{\phi} L_{lld}$
    \\
    \hspace{-0.2cm}6. \quad freeze $\psi_{dis}$ and $\phi$, compute DSD loss (\ref{dsd}) using $\{\textbf{M}_{i}^{s},\textit{y}_{i}^{s}\}$ 
    \\
    \hspace{-0.2cm}\quad \quad and $\{\textbf{M}_{j}^{t}\}$, then update $\theta_{bio}$, $\psi_{bio}$, $\theta_{dom}$ and $\psi_{dom}$:
    \\
    \hspace{-0.25cm}\qquad \qquad $\theta \leftarrow \theta -\gamma {{\nabla }_{\theta }}{{L}_{dsd}}$, \; $\theta \in \{{{\theta }_{bio}}, {{\psi}_{bio}}, {{\theta }_{dom}}, {{\psi}_{dom}}\}$
    \\
    \hspace{-0.2cm}7. \textbf{end for}
    \\
    \hspace{-0.2cm}8. \textbf{return} ${{\theta}_{bio}}$, ${{\psi }_{bio}}$
    \\
    \bottomrule[1.5pt]
  \end{tabular}
  \vspace{-1em}
\end{table}

\vspace{-0.5em}
\subsection{Mutual information minimization}
The last task strives to reduce the dependency between the biomarker embeddings and domain-related information via MIM. To extract domain-related information, a domain encoder $\theta_{dom}$ and a domain classifier $\psi_{dom}$ are utilized. The domain encoder takes in $\textbf{M}_{i}^{s}$ and $\textbf{M}_{j}^{t}$ to extract domain emebddings $\textbf{z}_{i}^{s}$ and $\textbf{z}_{j}^{t}$ respectively, which are used for domain prediction via the domain classifier. Therefore, $\theta_{dom}$ and $\psi_{dom}$ are jointly trained by minimizing the domain classification loss that is similar with (\ref{dis}) as:\begin{equation}
\begin{split}
{{L}_{dom}}=&\frac{1}{I}\sum\nolimits_{i=1}^{I}{\log f\left( \textbf{M}_{i}^{s};{{\theta }_{dom}},{{\psi }_{dom}} \right)}+ \\
&\frac{1}{J}\sum\nolimits_{j=1}^{J}{\log \left(1-f\left( \textbf{M}_{j}^{t};{{\theta }_{dom}},{{\psi }_{dom}} \right)\right)} \label{dom}
\end{split}
\end{equation}The embeddings $\textbf{z}_{i}^{s}$ and $\textbf{z}_{j}^{t}$ are domain-dependent and can be used to represent domain-related information.

\begin{table*}[t]
  \caption{Within-domain DSD results (\%) for different methods and corpora, where training and testing are performed in the target domain with labelled data, DSD systems are trained with 10 rounds, and mean $\pm$ standard-variance are reported.}
  \label{ind}
  \vspace{-1em}
  \centering
  \begin{tabular}{c|c|c|c|c|c|c}
    \thickhline
    \multirow{2}{*}{Methods} & \multicolumn{3}{|c}{UASPEECH} & \multicolumn{3}{|c}{TORGO} \\
    \cline{2-7}
    & WAR-utterance & UAR-utterance & ACC-speaker & WAR-utterance & UAR-utterance & ACC-speaker
    \\
    \hline
    RCNN \cite{korzekwa2019interpretable} & 85.71 $\pm$ 1.43 & 85.34 $\pm$ 1.50 & 93.57 $\pm$ 2.67 & 52.93 $\pm$ 3.78 & 54.25 $\pm$ 2.40 & 62.86 $\pm$ 2.86 \\
    RNN-A \cite{millet2019learning} & 86.78 $\pm$ 1.56 & 86.77 $\pm$ 1.64 & 94.29 $\pm$ 3.64 & 58.15 $\pm$ 1.83 & 58.02 $\pm$ 1.28 & 70.00 $\pm$ 5.35 \\
    CBRNN-A (proposed) & \textbf{87.87 $\pm$ 1.53} & \textbf{87.89 $\pm$ 1.56} & \textbf{95.71 $\pm$ 1.43} & \textbf{63.18 $\pm$ 1.14} & \textbf{62.76 $\pm$ 1.93} & \textbf{78.57 $\pm$ 5.82} \\
    \thickhline
  \end{tabular}
  \vspace{-1.5em}
\end{table*}

Then the mutual information between the biomarker embeddings \textbf{x} ($\textbf{x}_{i}^{s}$ or $\textbf{x}_{j}^{t}$) and domain embeddings \textbf{z} ($\textbf{z}_{i}^{s}$ or $\textbf{z}_{j}^{t}$) is used to measure the dependency as Kullback-Leibler (KL) divergence between their joint and marginal distribution: $\mathbf{H}(\mathbf{x},\mathbf{z})\!=\!{{D}_{KL}}(p(\mathbf{x},\mathbf{z});p(\mathbf{x})p(\mathbf{z}))$. As the computation of mutual information is challenging for high-dimensional continuous variables with unknown probability distributions, variational contrastive log-ratio upper bound (vCLUB) \cite{cheng2020club} is used to calculate the mutual information loss as:\begin{equation}
\begin{split}
   \textit{L}_{mi}\!=\!{{E}_{p(\mathbf{x},\mathbf{z})}}\!\left[ \log {{q}_{\phi }}\left( \mathbf{x}|\mathbf{z} \right) \right]\!-\!{{E}_{p(\mathbf{x})}}{{E}_{p(\mathbf{z})}}\left[ \log {{q}_{\phi }}\!\left( \mathbf{x}|\mathbf{z} \right) \right]
 \label{vclub}
\end{split}
\end{equation}where ${{q}_{\phi }}(\textbf{x}|\textbf{z})$ is the variational approximation of the ground-truth posterior of \textbf{x} given \textbf{z} and can be parameterized by a network $\phi$. During training, $\theta_{bio}$ and $\theta_{dom}$ are optimized to minimize $\textit{L}_{mi}$ (\ref{vclub}), while the variational approximation network $\phi$ is optimized to maximize the log-likelihood:\begin{equation}
    {{L}_{lld}}=\log {{q}_{\phi }}\left( \mathbf{x}|\mathbf{z} \right) \label{lld}
\end{equation}

\subsection{Integrating the learning tasks}
By combining the three learning tasks, the total DSD training loss used for updating $\theta_{bio}$, $\psi_{bio}$, $\theta_{dom}$ and $\psi_{dom}$ is:\begin{equation}
    {{L}_{dsd}}={{\lambda }_{1}}{{L}_{bio}}-{{\lambda }_{\text{2}}}{{L}_{dis}}+{{\lambda }_{3}}{{L}_{dom}}+{{\lambda }_{4}}{{L}_{mi}} \label{dsd}
\end{equation}where $\lambda_{k}$ (\textit{k}=1, 2, 3, 4) are positive constant weights. The final training process is summarized in Algorithm 1, where the well-trained biomarker encoder $\theta_{bio}$ and dysarthria classifier $\psi_{bio}$ are retained to perform the detection of dysarthria.

\section{Experiments}

\subsection{Experimental setup}
To verify the effectiveness of proposed methods, the UASPEECH \cite{kim2008dysarthric} and TORGO \cite{rudzicz2012torgo} corpora are used for experimentation. UASPEECH contains 15 dysarthric speakers (11 males and 4 females) with cerebral palsy, and 13 healthy speakers (9 males and 4 females). Each speaker has three blocks of utterances with isolated words, where the speech stimuli in each block contain 100 uncommon words and 155 repetitive words (i.e., 10 digits, 26 alphabets, 19 computer commands and 100 common words), which are recorded by 7-channel microphone arrays, we select the data of M6-channel for experiments. TORGO contains 7 dysarthric speakers (4 males and 3 females) with cerebral palsy or amyotrophic lateral sclerosis and 7 healthy speakers (4 males and 3 females). Different from UASPEECH, the speech stimuli include not only words, but also non-words and sentences. Words are mainly chosen from the word intelligibility section of the Frenchay Dysarthria Assessment \cite{enderby1980frenchay} and Yorkston-Beukelman Assessment \cite{yorkston1984assessment}. Non-words involve 5–10 repetitions of /iy-p-ah/, /ah-p-iy/, and /p-ah-t-ah-k-ah/ along with high-pitch and low-pitch vowels. Sentences are formed by Grandfather passage from the Nemours database \cite{menendez1996nemours}, 162 sentences from the sentence intelligibility section of Yorkston-Beukelman Assessment, 460 sentences of MOCHA database \cite{wrench2000multichannel} and spontaneously elicited descriptive texts. Due to discrepancies of speech stimuli types, phonetic context, articulation patterns, recording environments and devices, UASPEECH and TORGO can be treated as \textit{two different domains} with distinct data distributions. 

All speech signals are sampled at 16kHz, 80-band mel-spectrogram is calculated with hanning window of 25ms and hop length of 10ms. Utterance-level \textit{z}-score normalization for mel-spectrograms is performed before feeding them into the DSD system. The biomarker encoder and domain encoder adopt the same architecture that contains Convolution Banks and Recurrent Neural Network \cite{donahue2015long} with Attention \cite{hsiao2018effective} (CBRNN-A). There are 8 convolution banks with kernel size varying from 1 to 8, one-layer long-shot term memory (LSTM) with 128 units is employed for Recurrent Neural Network, attention module includes two linear layers (100 and 1 unit) with a softmax layer to obtain a vector that is used to weight the linear combination of LSTM outputs to form the final biomarker embedding and domain embedding. Both dysarthria and domain classifiers contain a linear layer with the sigmoid function, and the domain discriminator contains two linear layers with hidden size of 128 and 1. The variational approximation ${q_{\phi}}(\textbf{x}|\textbf{z})$ in (\ref{vclub}) is parameterized by the Gaussian distribution as ${q_{\phi}}(\textbf{x}|\textbf{z})=\mathcal{N} (\textbf{x}|\mu (\textbf{z}),diag({{\sigma }^{2}}(\textbf{z})))$ with mean $\mu (\textbf{z})$ and variance ${\sigma }^{2}(\textbf{z})$ inferred by two-way linear layers with a hidden size of 256. All networks are trained by the Adam optimizer \cite{kingma2014adam} for 8 epochs with learning rate $\alpha$, $\beta$ and $\gamma$ set to 1e-4, 1e-4 and 1e-3 respectively, and the weights $\lambda_1$, $\lambda_2$, $\lambda_3$ and $\lambda_4$ in loss (\ref{dsd}) are set to 1, 1e-1, 1 and 1e-4 respectively.

We compare CBRNN-A with other networks that also use mel-spectrograms to detect dysarthria, including Recurrent Convolutional Neural Network (RCNN) \cite{korzekwa2019interpretable} and Recurrent Neural Network with Attention (RNN-A) \cite{millet2019learning}. We adopt the leave-one-subject-out cross validation scheme, i.e., all speakers are used for training except the one that is left out for testing. Three evaluation metrics are used: (1) Utterance-level weighted average recall (WAR), denoting the ratio of utterances that are correctly classified; (2) Utterance-level unweighted average recall (UAR), denoting the accuracy per class averaged by total number of classes; (3) Speaker-level accuracy (ACC), denoting the ratio of speakers for which more than 50\% of the individual’s utterances are classified correctly.  

\begin{table*}[t]
  \caption{Cross-domain DSD results (\%) for different methods under different domain mismatch conditions.}
  \label{ood}
  \vspace{-1em}
  \centering
  \begin{tabular}{c|c|c|c|c|c|c}
    \thickhline
    \multirow{2}{*}{Methods} & \multicolumn{3}{|c}{UASPEECH $\rightarrow$ TORGO} & \multicolumn{3}{|c}{TORGO $\rightarrow$ UASPEECH} \\
    \cline{2-7}
    & WAR-utterance & UAR-utterance & ACC-speaker & WAR-utterance & UAR-utterance & ACC-speaker
    \\
    \hline
    RCNN \cite{korzekwa2019interpretable} & 32.73 $\pm$ 0.26 & 49.72 $\pm$ 0.21 & 50.00 $\pm$ 0.00 & 59.58 $\pm$ 1.55 & 59.66 $\pm$ 1.44 & 64.43 $\pm$ 4.84 \\
    +DAT \& MIM & 50.55 $\pm$ 8.75 & 58.06 $\pm$ 1.23 & 58.57 $\pm$ 9.48 & 64.41 $\pm$ 2.88 & 65.35 $\pm$ 2.82 & 67.14 $\pm$ 4.25 \\
    \hline
    RNN-A \cite{millet2019learning} & 34.82 $\pm$ 1.65 & 51.24 $\pm$ 0.94 & 50.00 $\pm$ 0.00 & 64.55 $\pm$ 1.82 & 64.48 $\pm$ 1.73 & 72.29 $\pm$ 3.93 \\
    +DAT \& MIM & 52.58 $\pm$ 4.78 & 57.46 $\pm$ 4.20 & 65.71 $\pm$ 5.35 & 67.20 $\pm$ 1.34 & 67.87 $\pm$ 1.29 & 75.00 $\pm$ 3.19 \\
    \hline
    CBRNN-A (proposed) & 35.21 $\pm$ 2.93 & 51.32 $\pm$ 1.27 & 50.00 $\pm$ 0.00 & 63.14 $\pm$ 1.95 & 62.71 $\pm$ 2.10 & 70.71 $\pm$ 5.95 \\
    +DAT & 53.68 $\pm$ 7.08 & 57.84 $\pm$ 6.38 & 62.86 $\pm$ 5.43 & 66.00 $\pm$ 1.89 & 66.47 $\pm$ 1.98 & 76.43 $\pm$ 1.75 \\
    +MIM & 43.80 $\pm$ 3.53 & 54.27 $\pm$ 3.63 & 51.43 $\pm$ 2.86 & 64.96 $\pm$ 3.08 & 65.15 $\pm$ 3.08 & 71.43 $\pm$ 5.31 \\
    +DAT \& MIM & \textbf{57.42 $\pm$ 4.74} & \textbf{60.70 $\pm$ 5.22} & \textbf{70.00 $\pm$ 5.29} & \textbf{68.44 $\pm$ 3.23} & \textbf{68.89 $\pm$ 3.26} & \textbf{79.29 $\pm$ 4.17} \\
    \thickhline
  \end{tabular}
  \vspace{-1.5em}
\end{table*}

\begin{table}[t]
  \caption{Utterance-level WAR (\%) for words, non-words and sentences under the mismatched condition of `UASPEECH $\rightarrow$ TORGO'.}
  \label{wns}
  \vspace{-1em}
  \centering
  \scalebox{0.95}{
  \begin{tabular}{c|c|c|c}
    \thickhline
    Methods & Words & Non-words & Sentences
    \\
    \hline
    CBRNN-A  & 32.43 $\pm$ 4.10 & 42.86 $\pm$ 1.27 & 29.88 $\pm$ 2.79 \\
    +DAT & 50.73 $\pm$ 5.60 & 48.19 $\pm$ 3.20 & 58.73 $\pm$ 1.52 \\
    +MIM & 40.89 $\pm$ 3.17 & 45.84 $\pm$ 2.15 & 44.12 $\pm$ 5.41 \\
    +DAT \& MIM & \textbf{55.10 $\pm$ 5.44} & \textbf{49.87 $\pm$ 1.43} & \textbf{64.05 $\pm$ 1.29} \\
    \thickhline
  \end{tabular}}
  \vspace{-1.6em}
\end{table}

\vspace{-0.5em}
\subsection{Experimental results and analysis}
\subsubsection{Within-domain DSD performance}
We first evaluate within-domain DSD performance, where the training and testing are both performed in a target domain, assuming that labelled data is provided, i.e., the ideal condition. The results are shown in Table \ref{ind}, we can see that the proposed CBRNN-A outperforms RCNN and RNN-A with higher utterance-level WAR, UAR and speaker-level ACC for both UASPEECH and TORGO corpora. This shows the effectiveness of CBRNN-A by using multiple convolution banks with varied kernel size to capture articulation patterns at different scales for accurate DSD.

\vspace{-0.5em}
\subsubsection{Cross-domain DSD performance}
Next, we consider two domain mismatch conditions: `UASPEECH $\rightarrow$ TORGO' and `TORGO $\rightarrow$ UASPEECH', where the former treats UASPEECH as the source domain and TORGO as the target domain, and vice versa for the latter. Results are shown in Table \ref{ood}. First, for `UASPEECH $\rightarrow$ TORGO', the performance of all DSD systems without DAT and MIM drops significantly. As TORGO has data imbalance where the ratio between healthy and dysarthric utterances is around 2:1, and healthy utterances are often incorrectly classified, less than 50\% WAR-utterance and only 50\% ACC-speaker are achieved. This shows the susceptibility of DSD systems to domain mismatch issues. Second, detection accuracy can be improved by using DAT or MIM, and the combination of DAT and MIM can greatly boost DSD performance for different kinds of networks, where the proposed CBRNN-A outperforms RCNN and RNN-A when DAT and MIM are used, showing the effectiveness of proposed methods for learning dysarthria-discriminative and domain-invariant biomarker embeddings for robust dysarthria detection. Third, compared with `TORGO $\rightarrow$ UASPEECH', larger improvements can be achieved under `UASPEECH $\rightarrow$ TORGO' condition by using DAT and MIM, e.g., the absolute values of WAR-utterance and ACC-speaker are increased with 22.2\% and 20.0\% respectively by using CBRNN-A. As UASPEECH contains utterances with limited words, while TORGO contains richer words and unseen speech stimuli types including non-words and sentences, the DSD systems trained on UASPEECH generalize poorly to TORGO. This can be verified by the utternace-level WAR results for words, non-words and sentences as shown in Table \ref{wns}, CBRNN-A performs worst for sentences, followed by words and non-words. With the proposed DAT and MIM, 34.2\%, 22.7\% and 7.0\% absolute increase in WAR can be achieved on average for sentences, words and non-words respectively. As TORGO contains richer speech stimuli, DSD systems trained on TORGO have better generalization capability, which can be further enhanced by the proposed DAT and MIM approaches.

\begin{figure}[t]
  \centering
  \centerline{\includegraphics[width=8cm]{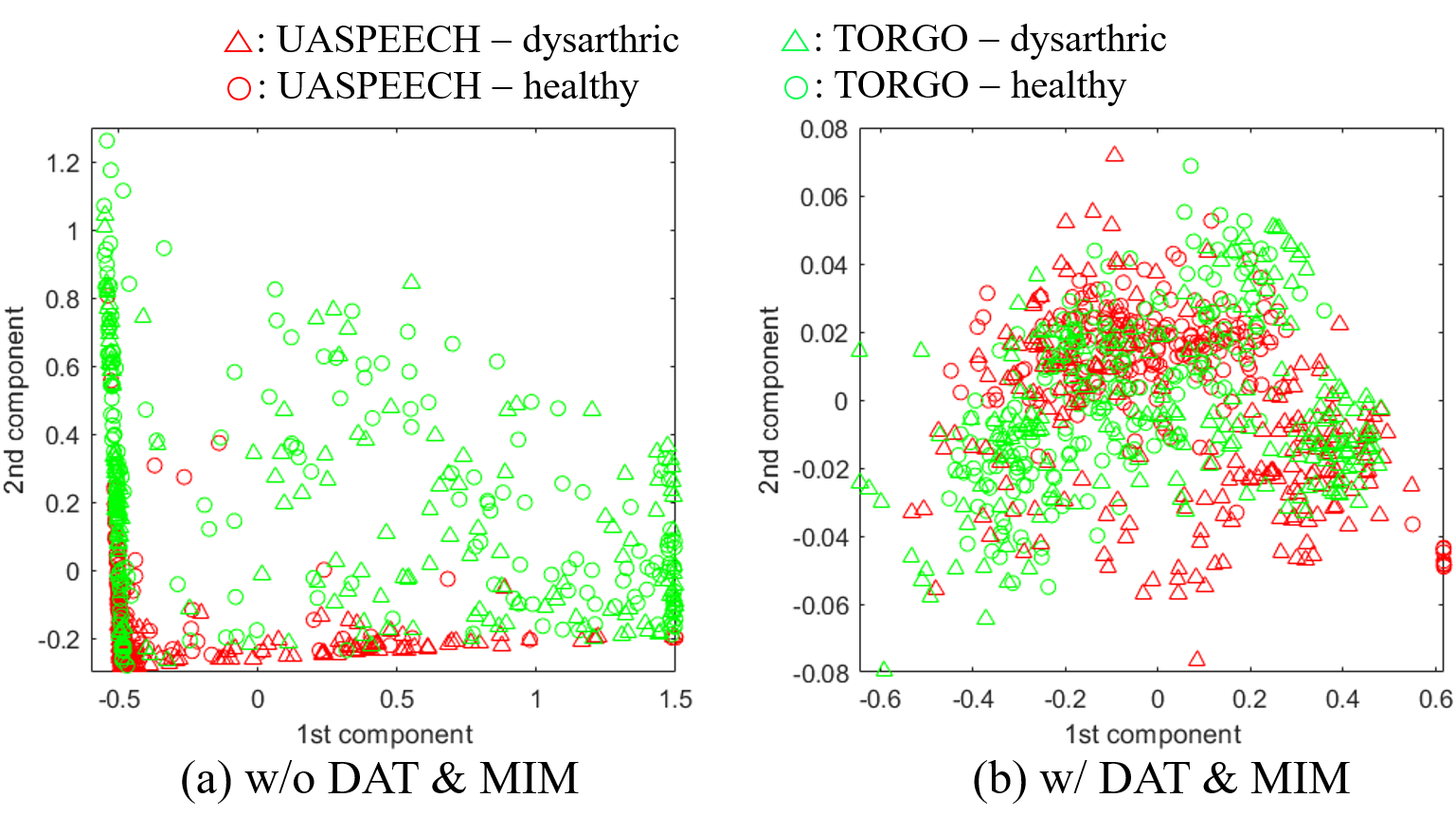}}
  \vspace{-1.2em}
  \caption{Visualization of the 1st and 2nd components of the biomarker embeddings extracted from utterances of UASPEECH and TORGO corpora based on DSD systems trained w/o DAT \& MIM (left) and w/ DAT \& MIM (right) under the mismatched condition of `UASPEECH $\rightarrow$ TORGO'.}\label{vis}
  \vspace{-1.8em}
\end{figure}

\vspace{-0.5em}
\subsubsection{Visualization of biomarker embeddings}
To acquire an intuition regarding how the UDA framework extracts effective biomarker embeddings, we consider `UASPEECH $\rightarrow$ TORGO' condition. Principal component analysis (PCA) is performed on biomarker embeddings extracted by the DSD systems trained without and with DAT \& MIM. The first and second components of PCA results are illustrated in Figure \ref{vis}, where different colors denote different domains, different shapes denote the presence or absence of dysarthria. We observe that the biomarker embeddings from UASPEECH and TORGO tend to be separate when DAT \& MIM is not used, while biomarker embeddings are mixed when DAT \& MIM is used, indicating that without additional regularizations, the biomarker embeddings contain domain cues, but these can be effectively removed by proposed DAT and MIM. Besides, it can be seen that dysarthric and healthy biomarker embeddings of DSD systems with DAT \& MIM are more dysarthria-discriminative with more obvious cluster formation than those of DSD systems without DAT \& MIM, which further proves the superiority of proposed methods to achieve higher detection accuracy across domains. 

\vspace{-0.5em}
\section{Conclusions}
This paper studies an under-explored field of DSD, i.e., cross-domain DSD where the DSD system is trained and tested on different domains with distinct data distributions. We propose a multi-task learning strategy, where the primary task performs DPC using labelled source data, while DAT and MIM tasks leverage large amounts of additional, unlabelled target-domain data that can be easily acquired to align the domain distributions. The proposed approach can obtain domain-invariant biomarker embeddings that contain critical indicators of dysarthria presence for accurate and robust detection. This is verified by extensive experiments with different kinds of network architectures. Our future study will focus on applying and improving the proposed UDA methods for more challenging domain mismatch conditions, e.g., cross-language condition.

\vspace{-0.5em}
\section{Acknowledgements}
This research is partially supported by the HKSARG Research Grants Council’s Theme-based Research Grant Scheme (Project No. T45- 407/19N).


\end{document}